\documentclass[largeformat]{interact}

\usepackage{pgfplots}
\pgfplotsset{compat=newest}
\pgfplotsset{plot coordinates/math parser=false}
\usepackage{tikz}

\usepackage[protrusion]{microtype}
\usepackage{amsmath,amssymb,amsthm}
\usepackage{mathtools}
\usepackage{mathrsfs}
\usepackage{nag}
\usepackage{xspace}
\usepackage{fancyhdr}
\usepackage[utf8]{inputenc}
\usepackage[backend=biber,style=apa,bibstyle=authoryear,citestyle=authoryear,sorting=nyt,dashed=false]{biblatex}
\let\cite\parencite 

\DeclareLanguageMapping{american}{american-apa}

\usepackage{soul}

\usepackage{xcolor}
\usepackage{nth}
\usepackage{aliascnt}
\usepackage{balance}\balance
\usepackage{enumitem}
\usepackage{mathtools}

\newtheorem{theorem}{Theorem}
\newaliascnt{cor}{theorem}
\newtheorem{cor}[cor]{Corollary}
\aliascntresetthe{cor}

\theoremstyle{definition}
\newtheorem{definition}{Definition}

\theoremstyle{remark}
\newtheorem{remark}{Remark}
\newtheorem{example}{Example}
\newtheorem{assumption}{Assumption}

\newcommand{\bR}{\ensuremath\mathbb{R}}
\newcommand{\bx}{\ensuremath\boldsymbol x}
\newcommand{\bu}{\ensuremath\boldsymbol u}
\newcommand{\by}{\ensuremath\boldsymbol y}
\newcommand{\cX}{\ensuremath\mathcal{X}}
\newcommand{\cU}{\ensuremath\mathcal{U}}
\newcommand{\cY}{\ensuremath\mathcal{Y}}
\newcommand{\cL}{\ensuremath\mathcal{L}}

\long\def\Note#1{\bgroup#1\egroup}

\makeatletter
\providecommand*{\diff}%
	{\@ifnextchar^{\DIfF}{\DIfF^{}}}
\def\DIfF^#1{%
	\mathop{\mathrm{\mathstrut d}}%
		\nolimits^{#1}\gobblespace}
\def\gobblespace{%
		\futurelet\diffarg\opspace}
\def\opspace{%
	\let\DiffSpace\!%
	\ifx\diffarg(%
		\let\DiffSpace\relax
	\else
		\ifx\diffarg[%
			\let\DiffSpace\relax
		\else
			\ifx\diffarg\{%
				\let\DiffSpace\relax
		\fi\fi\fi\DiffSpace}
\makeatother

\makeatletter
\newif\iftwocolumn%
\if@twocolumn\twocolumntrue\else\twocolumnfalse\fi%
\makeatother%

\begin{document}
\date{December 4, 2018}
\title{Passivity and Passivity Indices of Nonlinear Systems Under Operational Limitations using Approximations}
\def\shorttitle{Passivity Under Operational Limitations using Approximations}

\author{\name{Hasan Zakeri\thanks{CONTACT H. Zakeri. Email: hzakeri@nd.edu} and Panos J. Antsaklis}\affil{The authors are with the Department of Electrical Engineering, University of Notre Dame, Notre Dame, IN 46556 USA.}}

\maketitle
\begin{abstract}

In this paper, we will discuss how operational limitations affect input-output behaviours of the system. In particular, we will provide formulations for passivity and passivity indices of a nonlinear system given operational limitations on the input and state variables. This formulation is presented in the form of local passivity and indices. We will provide optimisation based formulation to derive passivity properties of the system through polynomial approximations. Two different approaches are taken to approximate the nonlinear dynamics of a system through polynomial functions; namely, Taylor's theorem and a multivariate generalisation of Bernstein polynomials. For each approach, conditions for stability, dissipativity, and passivity of a system, as well as methods to find its passivity indices, are given. Two different methods are also presented to reduce the size of the optimisation problem in Taylor's theorem approach. Examples are provided to show the applicability of the results.
\end{abstract}

\section{Introduction}\label{sec:intro}

Physical systems usually have inherent or imposed operational limitations. Whether it is a wall that limits the range of motion in a robot arm, or the limited force that we can apply to govern a system, these limitations should change our analysis of the system. In this paper, we will present how designers can consider knowledge of a system's operation in the input-output analysis of the system. In doing so, we focus on local dissipativity and extend it to local passivity and local passivity indices of a system given known operational limitations.

Passivity and dissipativity are fundamental concepts in control theory~\cite{willem.1,willem.2} and have been used in many applications~\cite{process.passivity,brogliato_dissipative_2007,constructive.nonlinear}. Traditionally, they are used to guarantee the stability of interconnected systems with robustness under parameter variations. Passivity can be seen as an abstraction of a system's behaviour, where the increase of energy stored in the system is less than or equal to the supplied energy. Passivity and dissipativity have shown great promise in the design of Cyber-Physical Systems (CPS)~\cite{passivity.cyber}. Their impact on CPS design comes from their compositional property in negative feedback and parallel (more generally, energy conserving) interconnections~\cite{process.passivity,hill.moylan.76,vanderschaft.gain.passivity.2017}. Besides, under mild assumptions, passivity implies stability (in the sense of \(\cL_2\) or asymptotic stability). A survey of applications of passivity indices in design of CPS can be found in~\cite{survey}.

Passivity and dissipativity have been treated the same way for linear systems and nonlinear systems; however, nonlinear systems require more detailed study. Specifically, we are interested in passivity/dissipativity behaviour of nonlinear systems under different operational conditions and different inputs. This distinction has always been made for local internal stability in the Lyapunov sense~\cite{sastry_nonlinear_2013,packard.roa.2009}, when the system is not externally excited. When there is an exogenous input applied to the system, few researchers have addressed the input-output behaviour of the system subject to operational constraints. In~\cite{topcu_local_2009}, the authors have addressed \(\mathcal L_2-\)gain of nonlinear systems locally, and extended the results to uncertain systems, both with unmodeled dynamics or those with parametric uncertainty. The present paper models the operational limitations as input and state constraints and focuses on local passivity and dissipativity of nonlinear systems.

Several attempts have been made to develop ways to find Lyapunov functionals for particular classes of systems, like linear systems or nonlinear systems described by polynomial fields~\cite{sos.lyapunov.2002}, However, a general methodology is still lacking. This paper addresses this gap by providing ways to find Lyapunov functionals through approximations, with an emphasis on dissipativity applications. 

The most common form of approximation is linearisation, which gives us a very tractable model with many analysis and synthesis tools available. The relation between passivity of a nonlinear system and passivity of its approximation is studied in~\cite{meng.linearization,meng.approximation}, where the authors show that when the linearised model is simultaneously strictly passive and strictly input passive, the nonlinear system is passive as well, within a neighbourhood of the equilibrium point around which the linearisation is done. However, in general, the linearisation is only valid within a limited neighbourhood, and the approximation error can be high. The relation between approximation error, the neighbourhood of study, and passivity/dissipativity are not evident in linearisation. {In~\cite{topcu_linearized_2009}, the relation between linearisation and optimisation based study of nonlinear systems is presented, and  conditions are presented based on linearisation for the feasibility of the optimisation problem.}

Here we propose approximations through multivariate polynomial functions. The methodology discussed in the present paper gives us approximate models in a well-defined neighbourhood of an operating point along with error bounds. Central to this approximation is \emph{the Stone-Weierstrass approximation theorem,} which states that under certain circumstances, any real-valued continuous function can be approximated by a polynomial function as closely as desired. Two different methods to approximate a nonlinear function have been employed here. The first methodology is Taylor's Theorem, which gives a polynomial approximation and bounds on the error function. Despite the simplicity and intuitiveness of the approximation, finding error bounds in this method requires complicated calculations, and results in large optimisation problems for real-world applications. The second approximation method is through Multivariate Bernstein Polynomials with more straightforward calculations that lead to a more tractable optimisation problem. Several results are given to test both local stability and local dissipativity of a nonlinear system through sum-of-squares optimisation and polynomial approximations of the system. Local QSR-dissipativity of the system, local passivity, and local passivity indices are also derived from the dissipativity results. {Both these methods require mild assumptions on the system and are generally applicable to broad classes of systems. The first approach requires a differentiability condition, which is satisfied by a majority of practical systems. The second approach only requires the Lipschitz condition to derive approximation bounds. This is not at all a limiting factor since the Lipschitz condition is essential in uniqueness and existence of solution~\cite{Khalil}.}

The organisation of this paper is as follows: \hyperref[sec:pre]{Section~\ref*{sec:pre}} presents introductory materials on dissipativity and passivity of dynamical systems. \hyperref[sec:local]{Section~\ref*{sec:local}} motivates the local passivity analysis of nonlinear systems under operational constraints through an example and introduces definitions for local dissipativity, passivity, and passivity indices. \hyperref[sec:local]{Section~\ref*{sec:approx}} presents two different approximations for a nonlinear system and methods for studying local dissipativity and passivity of the system through each approximation method. Specifically, the first part of \autoref{sec:approx} covers Taylor's theorem approach.  \autoref{th:taylor:dissipativity} gives conditions to check dissipativity of a nonlinear system with respect to a given supply rate function. However, the computational complexity of the optimisation can be quite high when the order of approximation or the order of the system's dynamics increase. \autoref{th:taylor:local:dissipativity:ellipsoid} reduces the size of the optimisation problem by approximating the error terms by ellipsoids providing optimisation constraints for specific admissible control and state space. \autoref{th:stabilityunforced2} formulates similar results for local stability of a nonlinear system.

The second part of \autoref{sec:approx} presents a generalisation of Berstein polynomials for multivariate functions followed by results on the analysis of a nonlinear system through its approximation by Bernstein polynomials. Specifically, \autoref{th:stabilityunforced3} presents conditions for local stability of a nonlinear system, while \autoref{th:bern:dissipativity} presents a method to check dissipativity of a system with respect to a given supply rate through Bernstein's approximation method. The rest of \autoref{sec:approx} presents conditions for \(QSR-\)dissipativity and passivity and methods to find passivity indices of a system through each approach. \hyperref[sec:ex]{Section~\ref*{sec:ex}} gives examples to demonstrate the applicability of the results. {Additional mathematical details of the Stone-Weierstrass theorem, Bernstein Polynomials, and generalised \(\mathcal S-\)procedure is in the appendix.} Finally, concluding remarks are given in \autoref{sec:conc}.

\section{Preliminaries}\label{sec:pre}

\subsection{Passivity and Dissipativity}
Consider a  continuous-time dynamical system \(\mathbf H:\bu\to\by\), where \(\bu\in\cU\subseteq\bR^m\) denotes the input and \(\by\in\cY\subseteq\bR^p\) denotes the corresponding output. Consider a real-valued function \(w(\bu(t),\by(t))\) (often referred as \(w(t)\) or \(w(\bu,\by)\) when clear from content) associated with \(\mathbf H\), called \emph{supply rate function.} We assume that \(w(t)\) satisfies 
\begin{equation}
	\int\limits_{t_0}^{t_1}\vert w(t)\vert\diff t<\infty,
\end{equation}
for every \(t_0\) and \(t_1.\) Now consider a continuous-time system described by
\begin{equation}\label{eq:system}
	\begin{aligned}
		\dot \bx&=f(\bx,\bu)\\
		\by&=h(\bx,\bu),
	\end{aligned}
\end{equation}
where \(f(\cdot,\cdot)\) and \(h(\cdot,\cdot)\) are Lipschitz mappings of proper dimensions, and assume the origin is an equilibrium point of the system; i.e., \(f(0,0)=0\) and \(h(0,0)=0.\) 
\begin{definition}\label{def:dissipativity}
	The system described by~\eqref{eq:system} is called \emph{dissipative with respect to supply rate function \(w(\bu(t),\by(t)),\)} if there exists a nonnegative function \(V(\bx),\) called the \emph{storage function,} such that \(V(0)=0\) and for all \(\bx_0\in\cX\subseteq\bR^n,\) all \(t_1\geq t_0,\) and all \(\bu\in\bR^m,\) we have
	\begin{equation}\label{eq:dissipativity}
		V(\bx(t_1))-V(\bx(t_0))\leq\int\limits_{t_0}^{t_1}w(\bu(t),\by(t))\diff t.
	\end{equation}
	where \(\bx(t_0)=x_0\) and \(\bx(t_1)\) is the state at \(t_1\) resulting from initial condition \(x_0\) and input function \(u(\cdot).\)
The inequality~\eqref{eq:dissipativity} is called \emph{dissipation inequality} and expresses the fact that the energy ``stored'' in the system at any time \(t\) is not more than the initially stored energy plus the total energy supplied to the system during this time. {If the dissipation inequality holds strictly, then the system~\eqref{eq:system} is called strictly dissipative with respect to supply rate function \(w(t).\)}
\end{definition}
If \(V(\bx)\) in \autoref{def:dissipativity} is differentiable, then~\eqref{eq:dissipativity} is equivalent to 
\begin{equation}\label{eq:dissipativitydiff}
	\dot V(\bx)\coloneqq\frac{\partial V}{\partial\bx}\cdot f(\bx,\bu)\leq w(u,y).
\end{equation}

According to the definition of supply rate, \(w(t)\) can take any form as long as it is locally integrable, however, we are particularly interested in the case when \(w(t)\) is quadratic in \(\bu\) and \(\by.\) More formally, a dynamical system is called \emph{\(QSR\)-dissipative} if its supply rate is given by
\begin{equation}\label{eq:QSR}
	w(\bu,\by)=\bu^\intercal R\bu+2\by^\intercal S\bu+\by^\intercal Q\by,
\end{equation}
where \(Q=Q^\intercal,\) \(S\) and \(R=R^\intercal\) are matrices of appropriate dimensions. One reason for considering such quadratic supply rate is that by selecting \(Q,S\) and \(R,\) we can obtain various notions of passivity and \(\cL_2\) stability. For instance, if a system is dissipative with supply rate given by~\eqref{eq:QSR} where \(R=\gamma^2I,S=0\) and \(Q=-I,\) then the system is \(\cL_2\) stable with finite gain \(\gamma>0\)~\cite{haddad_nonlinear_2008}.
\begin{definition}[Passivity~\cite{hill.moylan.76,willem.2}]\label{def:passive}
System~\eqref{eq:system} is called passive if it is dissipative with respect to the supply rate function \(w(\bu,\by)=\bu^\intercal\by.\)
\end{definition}
{The relation between different notions of passivity as well as their relation to Lyapunov stability and \(\cL_2\) stability has been extensively studied (see~\cite{meng.spr} and the references therein). }

\subsection{Passivity Indices}

The passivity index framework generalizes passivity to systems that may not be passive; In other words, it captures the level of passivity in  a system. If one of the systems in a negative feedback interconnection has ``shortage of passivity,'' it is possible that ``excess of passivity in the other system can assure the passivity or stability of the interconnection. More information on the compositional properties of passivity through passivity indices can be found in~\cite{process.passivity} and~\cite[p.~245]{Khalil}.
\begin{definition}[Input Feed-forward Passivity Index]
The system~\eqref{eq:system} is called \emph{input feed-forward passive (IFP)} if it is dissipative with respect to supply rate function \(w(\bu,\by)=\bu^\intercal\by-\nu\bu^\intercal\bu\) for some \(\nu\in\bR,\) denoted as IFP(\(\nu\)). Input feed-forward passivity (IFP) index for system~\eqref{eq:system} is the largest \(\nu\) for which the system is IFP. 
\end{definition}
\begin{definition}[Output Feedback Passivity]
The system~\eqref{eq:system} is called \emph{output feedback passive (OFP)} if it is dissipative with respect to supply rate function \(w(\bu,\by)=\bu^\intercal\by-\rho\by^\intercal\by\) for some \(\rho\in\bR,\) denoted as OFP(\(\rho\)). Output feedback passivity (OFP) index for system~\eqref{eq:system} is the largest \(\rho\) for which the system is OFP. 
\end{definition}

\section{Passivity Under Operational Limitations}\label{sec:local}

Unlike linear systems, important properties of nonlinear systems, like stability, are typically studied in a neighbourhood of an equilibrium point or other stationary sets and local analysis does not necessarily imply global stability. Local stability and region of convergence have been studied before using different techniques~\cite{henrion.roa,packard.roa.2009,packard.roa.2010}; however, dissipativity and passivity of nonlinear systems under constraints still require more in-depth study. 

To further expand this point, we start with an example~\cite{zakeri.acc2016}. Consider a nonlinear system governed by the following dynamics. 
  \begin{equation}\label{eq:motivationalexample:sys}
    \begin{aligned}
      \dot x&=-x+x^3+(-x+1)u\\
      y&= x-x^2+(\frac12x^2+1)u
    \end{aligned}
  \end{equation}
It is proved in~\cite{zakeri.acc2016} that this system is passive for 
  \begin{equation}
    \cX=\left\{x\mid x^2-1\leq0\right\}
  \end{equation}
  with a quartic storage function
  \begin{equation}
    V(x)=- 0.4581x^4 + 1.416x^2.
  \end{equation}
However, a closer look at the system's dynamics shows why it can not be globally passive.  This system has a stable equilibrium point at \(x=0.\) It also has two unstable equilibrium points at \(x=1\) and \(x=-1.\) The linearization of the  system around \(x=-1\) is \(\dot x=2x+u,y=3x+u,\) which is observable but not Lyapunov stable. Therefore, the nonlinear system~\eqref{eq:motivationalexample:sys} cannot be globally passive~\cite[Corollary~5.6]{haddad_nonlinear_2008}.
Furthermore, the passivity indices also depend on the operating region of the system. An example of this dependence is given in \cite{zakeri.acc2016}, where the system is proved to have an output passivity index of \(0.35\) for
\begin{equation}
    \cX=\{\bx\mid\Vert\bx\Vert_2\leq2.47\},
\end{equation}
but the index decreases as the state space radius increases, and at some point becomes negative and renders the system non-passive. Figure~\ref{fig:fig2} plots the provable OFP index of the system for different values of \(r\); where \(\cX=\{\bx\mid\Vert\bx\Vert_2\leq r\}.\) A simpler example can be found in~\cite[Chap.~2]{constructive.nonlinear}.
  \begin{figure}[b]
    \begin{center}
%
%
\definecolor{mycolor1}{rgb}{0.00000,0.44700,0.74100}%
\begin{tikzpicture}

\begin{axis}[%
width=0.7\columnwidth,
height=1.7in,
at={(0,0)},
scale only axis,
xmin=0,
xmax=20,
xlabel={$r$},
ymin=-0.1,
ymax=0.4,
ylabel={$\rho$},
title={$\rho$ versus $r$}
]
\addplot [color=mycolor1,solid,forget plot]
  table[row sep=crcr]{%
0.1	0.349999999923966\\
0.6	0.349999999913052\\
1.1	0.349999999767533\\
1.6	0.349999999929423\\
2.1	0.349999999799365\\
2.6	0.349999999789361\\
3.1	0.349999999696593\\
3.6	0.34999999988122\\
4.1	0.349999999545616\\
4.6	0.349999999724787\\
5.1	0.349999999764805\\
5.6	0.34999999973752\\
6.1	0.349985420138182\\
6.6	0.28256598257849\\
7.1	0.26962088433811\\
7.6	0.258453435180854\\
8.1	0.247922096161346\\
8.6	0.237641565929152\\
9.1	0.22741897500282\\
9.6	0.217135832835538\\
10.1	0.206708588424135\\
10.6	0.196071935503824\\
11.1	0.185170534969075\\
11.6	0.173954481769215\\
12.1	0.162376411979949\\
12.6	0.150389774672931\\
13.1	0.137947283458288\\
13.6	0.125000049495611\\
14.1	0.111496214469298\\
14.6	0.0973808822186584\\
15.1	0.0825937701195016\\
15.6	0.0670703474750098\\
16.1	0.0507385491951027\\
16.6	0.0335194690050002\\
17.1	0.0153243545072428\\
17.6	-0.00394505616536378\\
18.1	-0.0244002466408801\\
18.6	-0.0461657234171753\\
19.1	-0.0693837364972296\\
19.6	-0.094220282121114\\
};
\end{axis}
\end{tikzpicture}%
      \caption{OFP index \(\rho\) versus upper bound \(r\) on state norm~\cite{zakeri.acc2016}}\label{fig:fig2}
    \end{center}
  \end{figure}

Defining dissipativity properties for nonlinear systems with respect to constraints requires careful consideration of the admissible control and how we restrict the state space (operational limitations in this case are modeled as constraints over the input and state spaces). There are a few attempts in the literature to address this problem using different approaches. In~\cite{feedbackpassivity}, the authors defined local passivity in a neighbourhood of \(\bx=0,\bu=0\) with no further restriction. On the other hand, in~\cite{nijmeijer.passive}, local passivity is defined through a dissipation inequality holding for all \(\bx_0\in B_0\) and for all control \(\bu\) such that \(\Phi(t,\bx_0,\bu)\in B_0\) for \(t\geq0,\) where \(\Phi(t,\bx_0,\bu)\) is the full system response. In other words, local passivity is defined in a ball around the origin for the initial condition and for all inputs that do not drive the states ``away'' from the origin. While this assumption is useful, we are looking for a more explicit formulation of the admissible input space as well.  In~\cite{wstability}, local passivity is defined by putting constraints on the magnitude of the input signal and its derivative by using Sobolev spaces. This definition is based on suitable norms and inner products defined over the space. In~\cite{localdissipativity,potamoylan.interconnected}, local dissipativity is defined in terms of local internal stability regions and small gain inputs. {However, we are looking for an approach that can be naturally extended to passivity indices and has the same useful implications as passivity in the global sense.} Here We discuss local passivity indices, and we introduce approaches to determine these indices using polynomial approximations. To the best of our knowledge, local passivity indices for nonlinear systems were considered in~\cite{zakeri.acc2016} first.

\begin{definition}[Local Dissipativity]
  A given system of the form~\eqref{eq:system} is called locally dissipative if~\eqref{eq:dissipativity} holds for every \(\bu(t)\in\cU\subset\bR^m\) and \(\bx\in\cX\subset\bR^n,\) such that for every input signal \(\bu(t)\in\cU,\) the resulting state trajectories always remain in \(\cX.\) It is assumed that \(\cX\) contains the origin. 
\end{definition}
\begin{definition}\label{def:localpassive}
  A system is \emph{locally passive} if it is locally dissipative with respect to the supply rate function \(w(\bu,\by)=\bu^\intercal\by\) for every \(\bu(t)\in\cU\subset\bR^m\) and \(\bx\in\cX\subset\bR^n,\) such that for every input signal \(\bu(t)\in\cU,\) the state trajectories will always remain in \(\cX.\) 
\end{definition}
\begin{definition}
The local output feedback passivity (OFP) index is the largest gain that can be placed in positive feedback such that the interconnected system is passive and for every \(\bu(t)\in\cU\subset\bR^m,\) the state remains in \(\cX,\) i.e., \(\bx\in\cX\subset\bR^n\) for all times, where \(\cX\) and \(\cU\) satisfy the same assumptions as in~\autoref{def:passive}. This notion is equivalent to the following dissipative inequality holding for the largest \(\rho\), and for every \(\bu\in\cU\) and \(\bx\in\cX\)~\cite{zakeri.acc2016}
\begin{equation}\label{eq:ofp}
  \int_0^T\bu^\intercal\by\diff t\geq V(\bx(T))-V(\bx(0))+\rho\int_0^T\by^\intercal\by\diff t.
\end{equation}
\end{definition}
\begin{definition}
The local input feedforward passivity (IFP) index is the largest gain that can be put in a negative parallel interconnection with a system such that the interconnected system is passive and for every \(\bu(t)\in\cU\subset\bR^m,\) the state remains in \(\cX,\) i.e., \(\bx\in\cX\subset\bR^n\) for all times, where \(\cX\) and \(\cU\) satisfy the same assumptions as in~\autoref{def:passive}. This notion is equivalent to the following dissipative inequality holding for the largest \(\nu\), and for every \(\bu\in\cU\) and \(\bx\in\cX\) 
\begin{equation}\label{eq:ifp}
  \int_0^T\bu^\intercal\by\diff t\geq V(\bx(T))-V(\bx(0))+\nu\int_0^T\bu^\intercal\bu\diff t.
\end{equation}
A positive index indicates that the system has a positive feedforward path for all \(\bx\in\cX\) and that the zero dynamics are locally asymptotically stable. Otherwise, the index will be negative. 
\end{definition}
Local passivity and local dissipativity as defined in this section can offer many practical advantages. In most control applications, the aim is to keep the system working around an equilibrium, and given the practical limitations, global analysis is not always meaningful. For example, a pendulum, when it is upright, has very different behaviours than when it is hanging, even though both are equilibria of the system. This definition of local passivity and dissipativity addresses these kinds of operational conditions and actuator limitations. The same advantages that passivity and dissipativity have provided in the design and analysis of systems hold for local passivity and dissipativity as well. For example, if bounds on the signals are met, we will have the same compositional properties for local passivity as well; And this is not a limiting requirement, as most often the feedback loop is arranged to keep the signals within a desired region. This is a contrasting view to, for example, the notion of \emph{Equilibrium independent passivity,} where the dissipation inequality needs to hold against every possible equilibrium point~\cite{EIP.2011}. Equilibrium independent passivity is a generalisation of passivity to the cases where the exact location of the equilibrium point is unknown, mostly due to interconnection, uncertainty, and variation in parameters. On the other hand, local passivity enables us to have a more precise knowledge of the system within its operational conditions.

\section{Polynomial Approximations}\label{sec:approx}

Here, we will discuss methods to study certain behaviours of a system through its approximations. We will present two different methods of approximation along with related optimisation problems. First, recall a well-known theorem in approximation theory. 

\begin{theorem}[Weierstrass Approximation Theorem{~\cite{Apostol}}]
Suppose \(f(\cdot)\) is a real-valued and continuous function defined on the compact real interval \([a, b].\) Then for every \(\varepsilon> 0,\) there exists a polynomial \(p(x)\) (which might depend on \(\varepsilon\)) such that for all \(x \in [a, b],\) we have \(\vert f(x) - p(x)\vert < \varepsilon,\) or equivalently, the supremum norm \(\Vert f - p\Vert < \varepsilon.\)
\end{theorem}

This theorem was then generalised (by Marshall~H.~Stone) in two regards. First, it considers an arbitrary compact Hausdorff space \(X\) (here we take neighborhoods in \(\bR^n\)) instead of the real interval \([a,b]\). Second, it investigates a more general subalgebra (multivariate polynomials in \(\bR^n\) in this case), rather than the algebra of polynomial functions. This theorem is included in the appendix, but we will discuss direct results later on.

\subsection{Approach Based on Taylor's Theorem}
A direct result of the Stone-Weierstrass theorem is \emph{Taylor's theorem,} which gives a method of finding a polynomial approximation of a function and determining bounds on approximation error. The multivariate case of Taylor's theorem is reported in the Appendix.

To check local dissipativity of the system using Taylor's approximation, the dissipation inequality~\eqref{eq:dissipativity} needs to be rewritten by substituting \(f(\bx,\bu)\) with its Taylor approximation~\eqref{eq:taylortheorem}, and solved for every value of \(\bx\) and \(\bu\) in \(\cX\) and \(\cU.\) The remainder term is of course non-polynomial, and the exact value is not known. However, it can be bounded by~\eqref{eq:remainderbound}, so~\eqref{eq:dissipativity} holds for every value of \(R\) in those bounds. This is an infinite dimensional optimization problem, since \(\bx,\ \bu,\) and~\(R\) take infinite values. One way to deal with this problem is to bound \(R\) inside a polytope, by saying \(\underline r\leq R\leq\overline r,\) and rewrite the inequality for every vertex of this polytope. A similar approach is taken in~\cite{chesi.roa.2009}, for a simpler case where nonlinearity is only a function of one of the state variables and appears affinely in the dynamics. {This is not an efficient way to handle the uncertainty in \(R,\) since we need to solve the optimisation for all \(2^{n^2k}\) vertices of the polytope at the same time.} On the other hand, given the general structure of \(\cX\) and \(\cU,\) the same approach might not apply to take these bounds into account. 
Even when there is sparsity or other desirable properties in the problem, this is still a large problem to solve. To handle this problem one could use the generalised \(\mathcal S\)-Procedure to reduce the size of the program. These conditions should hold for a neighbourhood around the origin, and this fact should reflect in the formulation as well. The following theorems address these issues, but first, we will state the assumptions needed in the theorems.

\begin{assumption}\label{assum:1}
The input to system~\eqref{eq:system} is contained in \(\cU,\) i.e. \(\bu\in\cU,\) for all \(t\geq0\) and for every \(\bu\in\cU,\) the resulting trajectories of the system stay in \(\cX\) forever, i.e. \(\bx\in\cX,\) where \(\cX\) and \(\cU\) are defined appropriately. 
\end{assumption}

\begin{theorem}\label{th:taylor:dissipativity}
Consider the system defined in equation~\eqref{eq:system} that holds~\autoref{assum:1}. Also assume that \(f(\cdot,\cdot)\) and \(h(\cdot,\cdot)\) satisfy the assumptions for Taylor's Theorem. Define sets \(\cX\) and \(\cU\) as
\begin{align}
	\cX&=\left\{\bx\mid\bx(t)\in\bR^n\mid g_i(\bx(t))\leq0,i=1,\dots,I_X,\forall t\geq0\right\}\label{eq:cx}\\
	\cU&=\left\{\bu\mid\bu(t)\in\bR^m\mid g^\prime_j(\bu(t))\leq0,j=1,\dots,I_U,\forall t\geq0\right\}\label{eq:cu}.
\end{align}
This system is dissipative with respect to the polynomial supply rate function \(w(\bu,\by),\) if there exists a polynomial function \(V(\bx),\) called a storage function, that is the solution to the following optimization program
\iftwocolumn
\small
\begin{multline}\label{eq:taylor:dissipativity:inequality}
	V(\bx)+\sum_{i=1}^{I_X}s_{1;i}(\bx)g_i(\bx)\geq0\\
	-\sum_{i=1}^{n}\frac{\partial V(\bx)}{\partial x_i}\Bigg(
	\sum _{{\vert\alpha_1\vert+\vert\alpha_2\vert\mathrlap{\leq k-1}}}
	\frac{D^{\alpha_1}_{\bx} f_i(\bx,\bu)D^{\alpha_2 }_{\bu} f_i(\bx,\bu)}
	{\alpha_1!\alpha_2!}\bigg\vert_{\substack{\bx=0\\\bu=0}}\bx^{\alpha_1}\bu^{\alpha_2}\\
	+\sum_{|\beta_1|+|\beta_2|=k}r_{i;\beta_1,\beta_2}\bx^{\beta_1}\bu^{\beta_2}\Bigg)
		+w(\bu,\hat\by)\\
	-\sum_{i=1}^{n}\sum_{\vert\beta_1\vert+\vert\beta_2\vert=k}\bigg(s_{2;i,\beta_1,\beta_2}(\overline r_{i;\beta_1,\beta_2}-r_{i;\beta_1,\beta_2})\\
	+s_{3;i,\beta_1,\beta_2}(\overline r_{i;\beta_1,\beta_2}+r_{i;\beta_1,\beta_2})\bigg)\\
	-\sum_{i=1}^{p}\sum_{\vert\delta_1\vert+\vert\delta_2\vert=k}\bigg(s_{4;i,\delta_1,\delta_2}(\overline t_{i;\delta_1,\delta_2}-t_{i;\delta_1,\delta_2})\\
	+s_{5;i,\delta_1,\delta_2}(\overline t_{i;\delta_1,\delta_2}+t_{i;\delta_1,\delta_2})\bigg)\\
	+\sum_{i=1}^{I_X}s_{6;i}(\bx,\bu)g_i(\bx)+\sum_{j=1}^{I_U}s_{7;j}(\bx,\bu)g_j^\prime(\bu)\geq0
\end{multline}
\else
\begin{equation}\label{eq:taylor:dissipativity:inequality}
	\begin{gathered}
		V(\bx)+\sum_{i=1}^{I_X}s_{1;i}(\bx)g_i(\bx)\geq0\\
		-\sum_{i=1}^{n}\frac{\partial V(\bx)}{\partial x_i}\left(
		\!\sum _{{\vert\alpha_1\vert+\vert\alpha_2\vert\leq k-1}}\!\!\!
			\frac{D^{\alpha_1}_{\bx} f_i(\bx,\bu)D^{\alpha_2 }_{\bu} f_i(\bx,\bu)}
			{\alpha_1!\alpha_2!}\bigg\vert_{\substack{\bx=0\\\bu=0}}\bx^{\alpha_1}\bu^{\alpha_2}+\!\!\sum_{|\beta_1|+|\beta_2|=k}\!\!r_{i;\beta_1,\beta_2}\bx^{\beta_1}\bu^{\beta_2}\right)\\
			+w(\bu,\hat\by)\\
		-\sum_{i=1}^{n}\sum_{\vert\beta_1\vert+\vert\beta_2\vert=k}\left(s_{2;i,\beta_1,\beta_2}(\overline r_{i;\beta_1,\beta_2}-r_{i;\beta_1,\beta_2})+s_{3;i,\beta_1,\beta_2}(\overline r_{i;\beta_1,\beta_2}+r_{i;\beta_1,\beta_2})\right)\\
		-\sum_{i=1}^{p}\sum_{\vert\delta_1\vert+\vert\delta_2\vert=k}\left(s_{4;i,\delta_1,\delta_2}(\overline t_{i;\delta_1,\delta_2}-t_{i;\delta_1,\delta_2})+s_{5;i,\delta_1,\delta_2}(\overline t_{i;\delta_1,\delta_2}+t_{i;\delta_1,\delta_2})\right)\\
		+\sum_{i=1}^{I_X}s_{6;i}(\bx,\bu)g_i(\bx)+\sum_{j=1}^{I_U}s_{7;j}(\bx,\bu)g_j^\prime(\bu)\quad\geq0
	\end{gathered}
\end{equation}
\fi

for some nonnegative polynomials \(s_{1;i}\) to \(s_{7;i},\) where \(\hat\by=[\hat y_1,\dots,\hat y_p]^\intercal,\) and
\iftwocolumn
\begin{align}
\label{yhatj}
	\hat y_j=&\sum_{{\vert\gamma_1\vert+\vert\gamma_2\vert\leq k-1}}
			\frac{D^{\gamma_1}_{\bx} h_j(\bx,\bu)D^{\gamma_2 }_{\bu} h_j(\bx,\bu)}
			{\gamma_1!\gamma_2!} 
			\bx^{\gamma_1}\bu^{\gamma_2}\\
			&+\sum _{{\vert\delta_1\vert+\vert\delta_2\vert=k}}t_{j;\delta_1,\delta_2}\bx^{\delta_1}\bu^{\delta_2}.\notag
\end{align}
\else
\begin{equation}\label{yhatj}
\begin{aligned}
	\hat y_j&=\sum_{{\vert\gamma_1\vert+\vert\gamma_2\vert\leq k-1}}
			\frac{D^{\gamma_1}_{\bx} h_j(\bx,\bu)D^{\gamma_2 }_{\bu} h_j(\bx,\bu)}
			{\gamma_1!\gamma_2!} 
			\bx^{\gamma_1}\bu^{\gamma_2}
			&
			+
		\sum _{\mathclap{\vert\delta_1\vert+\vert\delta_2\vert=k}}t_{j;\delta_1,\delta_2}\bx^{\delta_1}\bu^{\delta_2}.
\end{aligned}
\end{equation}
\fi
Here, \(\overline r_{i;\beta_1,\beta_2}\) and \(\overline t_{i;\delta_1,\delta_2}\) are upper bounds for the remainder terms of Taylor's approximation of \(f(\cdot,\cdot)\) and \(h(\cdot,\cdot),\) respectively, which can be computed by~\eqref{eq:remainderbound}.

\begin{proof}
\autoref{th:Taylor} can be applied to approximate \(f(\bx,\bu)\) and \(h(\bx,\bu)\) in nonlinear system~\eqref{eq:system} as \(k\)-th order polynomials as follows 
\begin{equation}\label{eq:sys:approximate}
	\begin{aligned}
		\frac{\mathrm d\/x_i}{\mathrm d\/t}&=
		\sum _{{\vert\alpha_1\vert+\vert\alpha_2\vert\leq k-1}}
			\frac{D^{\alpha_1}_{\bx} f_i(\bx,\bu)D^{\alpha_2 }_{\bu} f_i(\bx,\bu)}
			{\alpha_1!\alpha_2!} 
			\bx^{\alpha_1}\bu^{\alpha_2}\\
			&+
		\sum _{\mathclap{\vert\beta_1\vert+\vert\beta_2\vert=k}}R_{i;\beta_1,\beta_2}(\bx,\bu )\bx^{\beta_1}\bu^{\beta_2},\\
		y_j&=\sum_{{\vert\gamma_1\vert+\vert\gamma_2\vert\leq k-1}}
			\frac{D^{\gamma_1}_{\bx} h_j(\bx,\bu)D^{\gamma_2 }_{\bu} h_j(\bx,\bu)}
			{\gamma_1!\gamma_2!} 
			\bx^{\gamma_1}\bu^{\gamma_2}\\
			&\qquad+
		\sum _{\mathclap{\vert\delta_1\vert+\vert\delta_2\vert=k}}T_{j;\delta_1,\delta_2}(\bx,\bu )\bx^{\delta_1}\bu^{\delta_2}.
	\end{aligned}
\end{equation}

We rewrite~\eqref{eq:dissipativity} and~\eqref{eq:dissipativitydiff} and substitute \(f(\cdot,\cdot)\) and \(h(\cdot,\cdot)\) with their Taylor's expansions~\eqref{eq:sys:approximate}. Since the remainders are not necessarily polynomial and their exact form are not known, we replace \(R_{i;\beta_1,\beta_2}(\bx,\bu)\) and \(T_{i;\delta_1,\delta_2}(\bx,\bu)\) with algebraic variables \(r_{i;\beta_1,\beta_2}\) and \(t_{i;\delta_1,\delta_2}\), whose bounds  can be written as
\begin{equation}\label{eq:errorbounds}
	\begin{aligned}
		-\overline r_{i;\beta_1,\beta_2}\leq r_{i;\beta_1,\beta_2}\leq\overline r_{i;\beta_1,\beta_2},\\
		-\overline t_{j;\delta_1,\delta_2}\leq t_{j;\delta_1,\delta_2}\leq\overline t_{j;\delta_1,\delta_2}.
	\end{aligned}
\end{equation}
Taking error bounds~\eqref{eq:errorbounds} and sets \(\cX\) and \(\cU\) defined in~\eqref{eq:cx} and~\eqref{eq:cu} and employing the generalised \(\mathcal S\)-Procedure to incorporate them with the dissipation inequality proves the theorem.
\end{proof}
\end{theorem}

Even though based on Taylor's theorem, the approximation can be as close as desired, there is always the problem of increasing the complexity as the size increases. More precisely, we will need \(4n^3k^2\) nonnegative polynomials as generalised $\mathcal S-$procedure multipliers for error bounds. If each of these multipliers is of degree \(\kappa,\) then the approximation will impose a total of approximately \(\kappa!(n+m)n^3k^2\) unknown variables to the optimisation problem. This increase in the size will become a problem even in the most straightforward examples; therefore it is necessary to derive a more tractable solution. The following theorem presents more tractable result by surrounding the approximation errors in an ellipsoid.

\begin{theorem}\label{th:taylor:local:dissipativity:ellipsoid}
Consider the system defined in equation~\eqref{eq:system} that holds \autoref{assum:1}. Also assume that \(f(\cdot,\cdot)\) and \(h(\cdot,\cdot)\) satisfy the assumptions of Taylor's Theorem, and sets \(\cX\) and \(\cU\) are defined as~\eqref{eq:cx} and~\eqref{eq:cu}, respectively. Then this system is locally dissipative with respect to the polynomial supply rate function \(w(\bu,\by),\) if there exists a polynomial \(V(\bx)\) called storage function that is solution to the following feasibility program
\iftwocolumn%
\small%
\begin{multline}\label{eq:taylor:local:dissipativity:ellipsoid:inequality}
	V(\bx)+\sum_{i=1}^{I_X}s_{1;i}(\bx)g_i(\bx)\geq0\\
	-\sum_{i=1}^{n}\frac{\partial V(\bx)}{\partial x_i}\left(\sum _{{\vert\alpha_1\vert+\vert\alpha_2\vert\mathrlap{\leq k-1}}}\frac{D^{\alpha_1}_{\bx} f_i(\bx,\bu)D^{\alpha_2 }_{\bu} f_i(\bx,\bu)}	{\alpha_1!\alpha_2!}\bigg\vert_{\substack{\bx=0\\\bu=0}}\bx^{\alpha_1}\bu^{\alpha_2}\right.\\
	\left.+\sum_{|\beta_1|+|\beta_2|=k}r_{i;\beta_1,\beta_2}\bx^{\beta_1}\bu^{\beta_2}\right)+w(\bu,\hat\by)\\
	-\sum_{i=1}^{n}s_{2;i}(\bx,\bu)\left(\overline r_{i}-\sum_{\vert\beta_1\vert+\vert\beta_2\vert=k}r_{i;\beta_1,\beta_2}^2\right)\\	
	-\sum_{i=1}^{p}s_{3;i}(\bx,\bu)\left(\overline t_{i}-\sum_{\vert\delta_1\vert+\vert\delta_2\vert=l}t_{i;\delta_1,\delta_2}^2\right)\\
	+\sum_{i=1}^{I_X}s_{4;i}(\bx,\bu)g_i(\bx)+\sum_{j=1}^{I_U}s_{5;j}(\bx,\bu)g_j^\prime(\bu)\geq0
\end{multline}
\else
\begin{equation}\label{eq:taylor:local:dissipativity:ellipsoid:inequality}
	\begin{gathered}
		V(\bx)+\sum_{i=1}^{I_X}s_{1;i}(\bx)g_i(\bx)\geq0\\
		-\sum_{i=1}^{n}\frac{\partial V(\bx)}{\partial x_i}\left(\!
		\sum _{{\vert\alpha_1\vert+\vert\alpha_2\vert\leq k-1}}\!\!
			\frac{D^{\alpha_1}_{\bx} f_i(\bx,\bu)D^{\alpha_2 }_{\bu} f_i(\bx,\bu)}
			{\alpha_1!\alpha_2!}\bigg\vert_{\substack{\bx=0\\\bu=0}}\bx^{\alpha_1}\bu^{\alpha_2}+\!\!\sum_{|\beta_1|+|\beta_2|=k}\!\!r_{i;\beta_1,\beta_2}\bx^{\beta_1}\bu^{\beta_2}\right)\\
			+w(\bu,\hat\by)\\
		-\sum_{i=1}^{n}s_{2;i}(\bx,\bu)\left(\overline r_{i}-\sum_{\vert\beta_1\vert+\vert\beta_2\vert=k}r_{i;\beta_1,\beta_2}^2\right)
		-\sum_{i=1}^{p}s_{3;i}(\bx,\bu)\left(\overline t_{i}-\sum_{\vert\delta_1\vert+\vert\delta_2\vert=l}t_{i;\delta_1,\delta_2}^2\right)\\
		+\sum_{i=1}^{I_X}s_{4;i}(\bx,\bu)g_i(\bx)+\sum_{j=1}^{I_U}s_{5;j}(\bx,\bu)g_j^\prime(\bu)\geq0
	\end{gathered}
\end{equation}	
\fi
for some nonnegative polynomials \(s_{1;i}\) to \(s_{5;i}\), where \(\hat\by=[\hat y_1,\dots,\hat y_p]^\intercal\), \(\hat y_j\) is defined as in~\eqref{yhatj}, \(\overline r_i\) and \(\overline t_i\) are defined as 
\begin{equation}
	\begin{aligned}
		\overline r_i&=\sum_{\vert\beta_1\vert+\vert\beta_2\vert=k}\overline r_{i;\beta_1,\beta_2}^2\\
		\overline t_i&=\sum_{\vert\delta_1\vert+\vert\delta_2\vert=k}\overline t_{i;\delta_1,\delta_2}^2.
	\end{aligned}
\end{equation}
\begin{proof}
Conditions in~\eqref{eq:taylor:local:dissipativity:ellipsoid:inequality} ensures, through generalised \(\mathcal S\)-Procedure, that 
\begin{align*}
	\sum_{\vert\beta_1\vert+\vert\beta_2\vert=k}r_{i;\beta_1,\beta_2}^2&\leq  
	\sum_{\vert\beta_1\vert+\vert\beta_2\vert=k}\overline r_{i;\beta_1,\beta_2}^2=\overline r_i\\
\intertext{and}
	\sum_{\vert\delta_1\vert+\vert\delta_2\vert=l}t_{i;\delta_1,\delta_2}^2&\leq
	\sum_{\vert\delta_1\vert+\vert\delta_2\vert=k}\overline t_{i;\delta_1,\delta_2}^2=\overline t_i
\end{align*}
which implies that the dissipation inequality holds for any value of \(r_{i;\beta_1,\beta_2}^2\) between \(-\overline r_{i;\beta_1,\beta_2}\) and \(\overline r_{i;\beta_1,\beta_2},\) and any value of \(t_{i;\delta_1,\delta_2}^2\) between \(-\overline t_{i;\delta_1,\delta_2}\) and \(\overline t_{i;\delta_1,\delta_2}.\) 
\end{proof}
\end{theorem}
The above program has only \(2n+2I_X+2I_U\) multipliers, where \(2n\) of these multipliers are for error bounds. This will yield to \(2\kappa!(n^2+mn)\) unknown variables in optimisation if each multiplier is of degree \(\kappa.\) This is a much smaller number compared to the former case.

\begin{remark}
If the order of approximation in either \autoref{th:taylor:dissipativity} or \autoref{th:taylor:local:dissipativity:ellipsoid} is 1, i.e. \(k=2,\) and the approximation error is negligible in the region of study, then the polynomial approximation will be  equivalent to linearization. Indeed, this is where optimisation and linearization based techniques coincide.  {Interested readers can refer to~\cite{topcu_linearized_2009} for more information on linearization based analysis versus optimisation based analysis of nonlinear systems. }
\end{remark}

Stability can also be studied through dissipativity results here.
\begin{cor}\label{th:stabilityunforced2}
The nonlinear system described by the following set of ordinary differential equations
\begin{equation}\label{eq:systemnou}
		\dot \bx=f(\bx)
\end{equation}
has a local stable equilibrium point at origin for \(\bx\in\cX,\) if there exist a polynomial \(V(\bx)\) and nonnegative polynomials \(s_{1;i},s_{2;i,\beta},s_{3;i,\beta}\) and  \(s_{4;i}\) for \(\vert\beta\vert=k\) satisfying the following  conditions
\iftwocolumn
\begin{gather}\notag
	V(\bx)-\phi_1(\bx)+\sum_{i=1}^{I_X}s_{1;i}(\bx)g_i(\bx)\geq0,\\\notag
	-\sum_{i=1}^{n}\frac{\partial V(\bx)}{\partial x_i}\left(
	\sum_{|\alpha |\leq k-1}{\frac {D^{\alpha }f_i(\bx)}{\alpha !}}\bigg\vert_{\bx=0}\bx^{\alpha }+\sum_{|\beta |=k}r_{i;\beta}\bx^\beta\right)\\\notag
	-\sum_{i=1}^{n}\sum_{\vert\beta\vert=k}\left(s_{2;i,\beta}(\overline r_{i\beta}-r_{i;\beta})+s_{3;i,\beta}(\overline r_{i;\beta}+r_{i;\beta})\right)\\
	+\sum_{i=1}^{I_X}s_{4;i}(\bx,\bu)g_i(\bx)
	-\phi_2(\bx)\geq0.\label{eq:stabilityunforced2}
\end{gather}
\else
\begin{equation}\label{eq:stabilityunforced2}
	\begin{gathered}
		V(\bx)-\phi_1(\bx)+\sum_{i=1}^{I_X}s_{1;i}(\bx)g_i(\bx)\geq0\\
		-\sum_{i=1}^{n}\frac{\partial V(\bx)}{\partial x_i}\left(
		\sum_{|\alpha |\leq k-1}{\frac {D^{\alpha }f_i(\bx)}{\alpha !}}\bigg\vert_{\bx=0}\bx^{\alpha }+\sum_{|\beta |=k}r_{i;\beta}\bx^\beta\right)\\
		-\sum_{i=1}^{n}\sum_{\vert\beta\vert=k}\left(s_{2;i,\beta}(\overline r_{i\beta}-r_{i;\beta})+s_{3;i,\beta}(\overline r_{i;\beta}+r_{i;\beta})\right)\\
		+\sum_{i=1}^{I_X}s_{4;i}(\bx,\bu)g_i(\bx)
		-\phi_2(\bx)\geq0
	\end{gathered}
\end{equation}
\fi
where \(\varphi_1\) and \(\varphi_2\) are arbitrary positive definite polynomials.
\end{cor}

\subsection{Approach Based on Bernstein Polynomials}\label{sec:approx:bernstein}
There is a second approach to Stone-Weierstrass theorem using \emph{Bernstein polynomials} used here to reduce the computation cost.  Details of Bernstein polynomials along with convergence proof and error margin can be found in the Appendix. 

The next theorem provides a numerical tool to test dissipativity of a nonlinear system through Bernstein polynomials approximations. {As mentioned before, \(QSR-\)dissipativity, passivity, and passivity indices can be derived from this theorem as well. Refer to \autoref{rem:qsr}, \autoref{rem:passive}, and \autoref{thm:index4} for details. }
\begin{theorem}\label{th:bern:dissipativity}
	The system defined in~\eqref{eq:system} is locally dissipative with respect to the supply rate function \(w(\bu,\by)\) over \(\cX\) and \(\cU\) defined as 
	\begin{align}
	\cX&=\left\{\bx\in\bR^n\mid \vert x_i\vert\leq\frac12,i=1,\dots,n\right\}\label{eq:cx:bernstein}\\
	\cU&=\left\{\bu\in\bR^m\mid \vert u_j\vert\leq\frac12,j=1,\dots,m\right\},\label{eq:cu:bernstein}
	\end{align}
	if there exist a polynomial function \(V(\bx)\) that is the solution to the following feasibility program
\iftwocolumn
	\begin{align}\notag
	V(\bx)&-\phi_1(\bx)+\sum_{i=1}^n\left(s_{1,i}(x_i-\frac12)-s_{2,i}(x_i+\frac12)\right)\geq0,\\\notag
	\sum_{i=1}^n\bigg(\!&-\frac{\partial V(\bx)}{\partial x_i}(b_i(\bx,\bu)+\varepsilon_i)
		+s_{3,i}(\varepsilon_i-\overline\varepsilon_i)-s_{4,i}(\varepsilon_i+\overline\varepsilon_i)\\
		\notag
	&+s_{5,i}(x_i-\frac12)-s_{6,i}(x_i+\frac12)	+s_{7,i}(u_i-\frac12)	\\\notag
	&-s_{8,i}(u_i+\frac12)+s_{9,i}(\varepsilon_i^\prime-\overline\varepsilon_i^\prime)-s_{10,i}(\varepsilon_i^\prime+\overline\varepsilon_i^\prime)\bigg)\\
	&+w(\bu,\boldsymbol b^\prime(\bx,\bu)+\boldsymbol \varepsilon^\prime) \geq0
	\label{eq:stability:unforced:conditions33}
	\end{align}
\else
	\begin{equation}\label{eq:stability:unforced:conditions33}
		\begin{gathered}
			V(\bx)-\phi_1(\bx)+\sum_{i=1}^n\left(s_{1,i}(x_i-\frac12)-s_{2,i}(x_i+\frac12)\right)\geq0,\\
			\sum_{i=1}^n\bigg(-\frac{\partial V(\bx)}{\partial x_i}(b_i(\bx,\bu)+\varepsilon_i)
				+w(\bu,\boldsymbol b^\prime(\bx,\bu)+\boldsymbol \varepsilon^\prime)
				+s_{3,i}(\varepsilon_i-\overline\varepsilon_i)-s_{4,i}(\varepsilon_i+\overline\varepsilon_i)\\
				+s_{5,i}(x_i-\frac12)-s_{6,i}(x_i+\frac12)
				+s_{7,i}(u_i-\frac12)-s_{8,i}(u_i+\frac12)
				+s_{9,i}(\varepsilon_i^\prime-\overline\varepsilon_i^\prime)-s_{10,i}(\varepsilon_i^\prime+\overline\varepsilon_i^\prime)\bigg)\\
				\geq0
		\end{gathered}	
	\end{equation}
\fi
	where 
\iftwocolumn\small
	\begin{multline}\label{eq:bi}
		b_i(\bx)=B_{\mu_1^i,\dots,\mu_n^i,\mu_{n+1}^i,\dots,\mu_{n+m}^i}(f)(x_1,\dots,x_n,u_1,\dots,u_m)=\\
			\sum_{\stackrel{0\le k_j\le \mu_j^i}{1\leq j\leq n+m}}\!
			f_i\left(\frac{k_1}{\mu_1^i}-\frac12,\dots,\frac{k_n}{\mu_n^i}-\frac12,\frac{k_{n+1}}{\mu_{n+1}^i}-\frac12,\dots,\frac{k_{n+m}}{\mu^i_{n+m}}-\frac12\right)\\
			\times\prod_{j=1}^n \left( \binom{\mu_j^i}{k_j} (x_j+\frac12)^{k_j} (\frac12-x_j)^{\mu_j^i-k_j} \right)\\
			\times\prod_{j=1}^m \left( \binom{\mu_{j+n}^i}{k_{j+n}} (u_j+\frac12)^{k_{j+n}} (\frac12-u_j)^{\mu_{j+n}^i-k_{j+n}} \right)
	\end{multline}
\else
	\begin{multline}\label{eq:bi}
		b_i(\bx)=B_{\mu_1^i,\dots,\mu_n^i,\mu_{n+1}^i,\dots,\mu_{n+m}^i}(f)(x_1,\dots,x_n,u_1,\dots,u_m)=\\
			\sum_{\stackrel{0\le k_j\le \mu_j^i}{1\leq j\leq n+m}}
			f_i\left(\frac{k_1}{\mu_1^i}-\frac12,\dots,\frac{k_n}{\mu_n^i}-\frac12,\frac{k_{n+1}}{\mu_{n+1}^i}-\frac12,\dots,\frac{k_{n+m}}{\mu^i_{n+m}}-\frac12\right)\\
			\times\prod_{j=1}^n \left( \binom{\mu_j^i}{k_j} (x_j+\frac12)^{k_j} (\frac12-x_j)^{\mu_j^i-k_j} \right)\\
			\times\prod_{j=1}^m \left( \binom{\mu_{j+n}^i}{k_{j+n}} (u_j+\frac12)^{k_{j+n}} (\frac12-u_j)^{\mu_{j+n}^i-k_{j+n}} \right)
	\end{multline}
\fi
	and \(\boldsymbol b^\prime(\bx,\bu)=[b_1^\prime,\dots,b_p^\prime]^\intercal,\) where
	\begin{multline}\label{eq:bprime}
		b^\prime_i(\bx)=B_{\eta_1^i,\dots,\eta_n^i,\eta_{n+1}^i,\dots,\eta_{n+m}^i}(h)(x_1,\dots,x_n,u_1,\dots,u_m)=\\
			\sum_{\stackrel{0\le k_j\le \eta_j^i}{1\leq j\leq n+m}}
			h_i\left(\frac{k_1}{\eta_1^i}-\frac12,\dots,\frac{k_n}{\eta_n^i}-\frac12,\frac{k_{n+1}}{\eta_{n+1}^i}-\frac12,\dots,\frac{k_{n+m}}{\eta^i_{n+m}}-\frac12\right)\\
			\times\prod_{j=1}^n \left( \binom{\eta_j^i}{k_j} (x_j+\frac12)^{k_j} (\frac12-x_j)^{\eta_j^i-k_j} \right)\\
			\times\prod_{j=1}^m \left( \binom{\eta_{j+n}^i}{k_{j+n}} (u_j+\frac12)^{k_{j+n}} (\frac12-u_j)^{\eta_{j+n}^i-k_{j+n}} \right).
	\end{multline}
\end{theorem}
\begin{remark}
The above theorem only imposes \(6n+2m+2p\) multipliers, which is a great improvement over Taylor's approach. The drawback here is that the latter is limited to \(\cX\) and \(\cU\) defined in~\eqref{eq:cx:bernstein} and~\eqref{eq:cu:bernstein}, therefore a scaling of variables is necessary if the region of study is different.
\end{remark}

The following theorem presents a stability test for a nonlinear system through a Bernstein approximation.
\begin{cor}\label{th:stabilityunforced3}
	The system described by~\eqref{eq:systemnou} is locally stable, if there exist a polynomial \(V(\bx)\) and nonnegative polynomials \(s_{j,i}\) for \(1\leq i\leq n\) and \(1\leq j\leq6\) that are solution to the following feasibility program.
\iftwocolumn\small
\begin{align}\notag
V(\bx)&-\phi_1(\bx)+\sum_{i=1}^n\left(s_{1,i}(x_i-\frac12)-s_{2,i}(x_i+\frac12)\right)\geq0\\
\notag
\sum_{i=1}^n\bigg(&-\frac{\partial V(\bx)}{\partial x_i}(b_i(\bx)+\varepsilon_i)
	+s_{3,i}(\varepsilon_i-\overline\varepsilon_i)-s_{4,i}(\varepsilon_i+\overline\varepsilon_i)\\
&+s_{5,i}(x_i-\frac12)-s_{6,i}(x_i+\frac12)\bigg)-\phi_2(\bx)\geq0
\label{eq:stability:unforced:conditions3}
\end{align}
\else
\begin{equation}\label{eq:stability:unforced:conditions3}
	\begin{gathered}
		V(\bx)-\phi_1(\bx)+\sum_{i=1}^n\left(s_{1,i}(x_i-\frac12)-s_{2,i}(x_i+\frac12)\right)\geq0\\
		\sum_{i=1}^n\left(-\frac{\partial V(\bx)}{\partial x_i}(b_i(\bx)+\varepsilon_i)
			+s_{3,i}(\varepsilon_i-\overline\varepsilon_i)-s_{4,i}(\varepsilon_i+\overline\varepsilon_i)
			+s_{5,i}(x_i-\frac12)-s_{6,i}(x_i+\frac12)\right)\\-\phi_2(\bx)\geq0
	\end{gathered}	
\end{equation}
\fi
where 
\iftwocolumn
\begin{multline}
	b_i(\bx)=B_{m_1^i,\dots,m_n^i}(f)(x_1,\dots,x_n)=\\
		\sum_{\stackrel{0\le k_j\le m_j^i}{1\leq j\leq n}}
		f_i\left(\frac{k_1}{m_1^i}-\frac12,\dots,\frac{k_n}{m_n^i}-\frac12\right)\times\\
		\prod_{j=1}^n \left( \binom{m_j^i}{k_j} (x_j+\frac12)^{k_j} (\frac12-x_j)^{m_j^i-k_j} \right)
\end{multline}
\else
\begin{multline}
	b_i(\bx)=B_{m_1^i,\dots,m_n^i}(x_1,\dots,x_n)=\\
		\sum_{\stackrel{0\le k_j\le m_j^i}{1\leq j\leq n}}
		f_i\left(\frac{k_1}{m_1^i}-\frac12,\dots,\frac{k_n}{m_n^i}-\frac12\right)
		\prod_{j=1}^n \left( \binom{m_j^i}{k_j} (x_j+\frac12)^{k_j} (\frac12-x_j)^{m_j^i-k_j} \right)
\end{multline}
\fi
is the Bernstein approximation of function \(f_i(\bx)\) in \(\bx\in[-\frac12,\frac12]^n,\) and \(\overline\varepsilon_i\) are bounds on approximation error which can be determined through~\eqref{eq:bernstein.errorbound}.
\end{cor}
\begin{remark}
The above theorem gives a local result for \(\bx\in[-\frac12,\frac12]^n.\) If a different region is meant to be studied, a scaling of state variables is necessary in advance. 
\end{remark}

\begin{remark}
In all of the theorems in this section, the supply rate is a polynomial function. This assumption is not limiting, and several control problems have a formulation as dissipation inequality form with a polynomial supply rate function (some are presented later on in this section, other examples are listed in~\cite{ebenbauer.dissipation}). However, if a non-polynomial function is desired, a similar approximation should be performed for the supply rate function as well. Such an approximation can be carried out similarly and will not be repeated here.
\end{remark}

{
\begin{remark}
Take note that the conditions on the theorems provided in this section are in the form of polynomial nonnegativity. This is a difficult problem to solve, even for simple cases, but the nonnegativity conditions can be relaxed into polynomial optimisation. The most popular way to relax the conditions is the use of sum of squares (SOS) programming, which converts the polynomial nonnegativity problem into a semidefinite optimisation program~\cite{sostutorial,lasserre_global_2001}. Novel approaches recently introduced in~\cite{ahmadi_dsos_2017} relax the conditions into linear programming and second-order cone programming, which are more efficient to solve. The examples in \autoref{sec:ex} are solved using SOSTOOLS~\cite{sostutorial}.
\end{remark}
}
\subsection{Passivity and Passivity Indices}\label{sec:dissipativity}
As mentioned in \autoref{sec:pre}, passivity is a special case of dissipativity, so we can study passivity and passivity indices of a system using either one of the approaches discussed earlier in this section. Here, for completeness, we state the results for {$QSR-$dissipativity,} passivity, and passivity indices. 


{
\begin{remark}\label{rem:qsr}
The nonlinear system defined in~\eqref{eq:system} is locally $QSR-$dissipative, if it is locally dissipative with respect to supply rate function
\begin{equation}\label{eq:supplyrate:qsr}
	w(\bu,\by)=\by^\intercal Q\by+2\by^\intercal S\bu+\bu^\intercal R\bu
\end{equation}
where \(Q,S,\) and \(R\) are constant matrices of appropriate dimension and \(Q\) and \(R\) are symmetric. This can be checked using any of the Theorems~\ref{th:taylor:dissipativity}, \ref{th:taylor:local:dissipativity:ellipsoid}, and~\ref{th:bern:dissipativity}.
\end{remark}
\begin{remark}\label{rem:passive}
The nonlinear system defined in~\eqref{eq:system} is locally passive, if it is locally dissipative with respect to supply rate function
\begin{equation}\label{eq:supplyrate:passive}
	w(\bu,\by)=\bu^\intercal\by.
\end{equation}
Local passivity of the system can be checked using Theorems~\ref{th:taylor:dissipativity},\ref{th:taylor:local:dissipativity:ellipsoid}, and~\ref{th:bern:dissipativity}. This system is called \emph{locally Input Feed-forward Output Feedback Passive (IF-OFP),} if it is locally dissipative with respect to the well-defined supply rate:
\begin{equation}
	w(\bu,\by)={\bu}^\intercal\by-\rho\by^\intercal\by-\nu\bu^\intercal\bu
\end{equation}
for some \(\nu,\rho\in\bR.\)
\end{remark}
}
The following two theorems present ways to find passivity indices of a system and can be easily derived from previous theorems and definitions.

\begin{theorem}\label{thm:index3}
	The nonlinear system~\eqref{eq:system} has local output feedback passivity (OFP) index of \(\rho,\) if conditions in \autoref{th:taylor:local:dissipativity:ellipsoid} hold for the largest \(\rho,\) where \(w(\bu,\by)\) is given as
	\begin{equation}\label{eq:supply:ofp}
		w(\bu,\by)=\bu^\intercal\by-\rho\by^\intercal\by.
	\end{equation}	
	\(\nu\) is local input feedforward passivity (IFP) for the system if it is the biggest number satisfying condition in \autoref{th:taylor:local:dissipativity:ellipsoid} with \(w(\bu,\by)\) defined as
	\begin{equation}\label{eq:supply:ifp}
		w(\bu,\by)=\bu^\intercal\by-\nu\bu^\intercal\bu.
	\end{equation}
	Here, local means for \(\bx\) and \(\bu\) belonging to \(\cX\) and \(\cU\) defined in~\eqref{eq:cx} and~\eqref{eq:cu}.
\end{theorem}
\begin{theorem}\label{thm:index4}
	The nonlinear system~\eqref{eq:system} has local OFP (IFP) index of \(\rho\) (\(\nu\)) for \(\cX\) and \(\cU\) defined in~\eqref{eq:cx:bernstein} and~\eqref{eq:cu:bernstein}, if \(\rho\) (\(\nu\)) is the largest value satisfying conditions in \autoref{th:bern:dissipativity}, with \(w(\bu,\by)\) defined in~\eqref{eq:supply:ofp} (or~\eqref{eq:supply:ifp}, respectively).
\end{theorem}

\section{Examples}\label{sec:ex}
Examples are provided here to demonstrate how to employ the given techniques to approximate a nonlinear system and to verify stability and passivity. \autoref{ex:stability} demonstrate the use of Taylor's approximation theorem and determining the stability of a dynamic system through \autoref{th:stabilityunforced2}. \autoref{ex:passivity} studies passivity of a nonlinear system using Taylor's approximation theorem as in \autoref{th:taylor:dissipativity}. \autoref{ex:pendulum} uses Bernstein polynomials to approximate the dynamics of a simple pendulum and demonstrates the use of \autoref{th:stabilityunforced3} as well. \autoref{ex:bernstein} shows the use of multivariable Bernstein polynomials and \autoref{th:bern:dissipativity}.

\begin{example}[Stability]\label{ex:stability}
    Consider the system as 
    \begin{equation}\label{eq:ex1:system}
        \begin{aligned}
            \dot x_1&=x_2,\\
            \dot x_2&=-2x_2-x_1\cos(x_1 + x_2);
        \end{aligned}
    \end{equation}
This system is nonlinear and non-polynomial. It is not trivial to find a Lyapunov functional to check stability or dissipativity of the system. Employing Lyapunov's indirect method will also not give us every detail about the system, including how close to the equilibrium we need to stay to remain stable, or what kind of inputs can keep the system dissipative.

Assume \(\bx=\begin{bmatrix}x_1&x_2\end{bmatrix}^\intercal\) and \(p(\bx)=x_1\cos(x_1+x_2).\) Using \autoref{th:Taylor} and~\eqref{eq:sys:approximate} we can rewrite \(p(\bx)\) as a \nth{6} order approximation plus remainder as follows.
\iftwocolumn
\begin{align}\notag
	p(\bx)={}&\sum_{i=0,j=0}^{i+j\leq6}\frac{\partial^i f(\bx)}{i!\partial x_1^i}\times\frac{\partial^jf(\bx)}{j!\partial x_2^j}x_1^ix_2^j+\sum_{i=0}^7R_i(\bx)x_1^ix_2^{(7-i)}\\ \notag
	={}&x_1^5/24 + (x_1^4x_2)/6 + (x_1^3x_2^2)/4 - x_1^3/2+ (x_1^2x_2^3)/6 \\
	&{-} x_1^2x_2 + (x_1x_2^4)/24 - (x_1x_2^2)/2 + x_1\\
	&+\sum_{i=0}^7R_i(\bx)x_1^ix_2^{(7-i)}.\notag
\end{align}
\else
\begin{equation}
	\begin{aligned}
		p(\bx)={}&\sum_{i=0,j=0}^{i+j\leq6}\frac{\partial^i f(\bx)}{i!\partial x_1^i}\cdot\frac{\partial^jf(\bx)}{j!\partial x_2^j}x_1^ix_2^j+\sum_{i=0}^7R_i(\bx)x_1^ix_2^{(7-i)}\\
		={}&x_1^5/24 + (x_1^4x_2)/6 + (x_1^3x_2^2)/4 - x_1^3/2+ (x_1^2x_2^3)/6 \\&{-} x_1^2x_2 + (x_1x_2^4)/24 - (x_1x_2^2)/2 + x_1+\sum_{i=0}^7R_i(\bx)x_1^ix_2^{(7-i)}.
	\end{aligned}
\end{equation}
\fi
However, the functions \(R_i\) are not polynomial, so we bound them based on~\eqref{eq:remainderbound} as
\iftwocolumn
\begin{align}\notag
	|R_0|&\leq2.0\times10^{-4}&&|R_1|\leq0.0028&&|R_2|\leq0.0125\\
	|R_3|&\leq0.0279&&|R_4|\leq0.0349&&|R_5|\leq0.0252\\\notag
	|R_6|&\leq0.0097&&|R_7|\leq0.0016
\end{align}
\else
\begin{equation}
	\begin{aligned}
		|R_0|&\leq2.0\times10^{-4}&&|R_1|\leq0.0028&&|R_2|\leq0.0125&&|R_3|\leq0.0279\\
		|R_4|&\leq0.0349&&|R_5|\leq0.0252&&|R_6|\leq0.0097&&|R_7|\leq0.0016
	\end{aligned}
\end{equation}
\fi
for \(|x_1|\leq1,|x_2|\leq1.\) Applying \autoref{th:stabilityunforced2} to above approximation will prove that the origin is a stable equilibrium point for the system for \(|x_1|\leq1,|x_2|\leq1.\) 
Stability is proved by a quartic Lyapunov functional
\iftwocolumn
\begin{equation}\begin{aligned}
V_1(\bx)=&	- 39.73x_1^4 + 1204.0x_1^3x_2  + 99.79x_1^2x_2^2\\ 
 		&- 106.1x_1^2 + 748.7x_1x_2^3   + 0.0002435x_2^4 
\end{aligned}\end{equation}
\else
\begin{equation}
V_1(\bx)=	- 39.73x_1^4 + 1204.0x_1^3x_2  + 99.79x_1^2x_2^2  - 106.1x_1^2 + 748.7x_1x_2^3   + 0.0002435x_2^4 
\end{equation}
\fi
Note that the function \(V_1(\bx)\) is not positive (semi)definite, but it is nonnegative for  \(|x_1|\leq1,|x_2|\leq1.\) 
\end{example} 
\begin{example}[Passivity]\label{ex:passivity}
	Now consider the system
	\begin{equation}\label{eq:ex2:system}
		\begin{aligned}
			\dot x_1&=x_2,\\
			\dot x_2&=-2x_2-x_1\cos(x_1 + x_2)+u;\\
			y&=x_2\\
		\end{aligned}
	\end{equation}
By approximating this system using \autoref{th:Taylor}, we can prove that the system is passive with the following storage function
\iftwocolumn
\begin{align}\notag
	V(\bx)=&- 23.63x_1^4 
		+ 674.4x_1^3x_2 
		+ 58.66x_1^2x_2^2 
		- 62.39x_1^2 \\
		&+ 422.4x_1x_2^3 
		- 4.08\times10^{-4}x_2^4 
\end{align}
\else
\begin{equation}
	V(\bx)=- 23.63x_1^4 
		+ 674.4x_1^3x_2 
		+ 58.66x_1^2x_2^2 
		- 62.39x_1^2 
		+ 422.4x_1x_2^3 
		- 4.08\times10^{-4}x_2^4 
\end{equation}
\fi
\end{example}
\begin{example}[Simple Pendulum]\label{ex:pendulum}
The equations of motion for a simple pendulum are given as
\begin{equation}
	\begin{aligned}
		\dot\theta&=\omega,\\
		\dot\omega&=-\sin\theta-\omega.
	\end{aligned}
\end{equation}
Here, we will use the approach based on Bernstein Polynomilas to study this system. Assuming bounds on states as \(\vert\theta\vert\leq0.5,\vert\omega\vert\leq0.5\) and change of variables as
\begin{equation}
	x_1=\theta+\tfrac12,\quad x_2=\omega+\tfrac12
\end{equation}
result in the following dynamical equation
\begin{equation}
	\begin{aligned}
		\dot x_1&=x_2-\tfrac12,\\
		\dot x_2&=-\sin(x_1-\tfrac12)-x_2+\tfrac12.
	\end{aligned}
\end{equation}
A \nth{6}-order approaximation of this system based on Bernstein approach can be derived as
{\small\begin{align}\notag
	\dot x_1=&x_2-\tfrac12,\\
	\dot x_2= &8.9\times10^{-16}x_1^6 - 7.6\times10^{-4}x_1^5 + 1.9\times10^{-3}x_1^4\\
			 &+ 0.089x_1^3 - 0.14x_1^2 - 0.91x_1 + 0.48 +\varepsilon-x_2+\tfrac12.\notag
\end{align}}
where \(\vert\varepsilon\vert\leq0.04\) is the approximation error. Assuming \(u=0,\) \autoref{th:stabilityunforced3} proves that the system is locally stable based on the following Lyapunov function:
{\small\begin{align*}
\iftwocolumn
	V=&-1.49\omega^6 + 2.45\omega^5\theta + 13.62\omega^4\theta^2 + 37.74\omega^3\theta^3 - 3.67\omega^2\theta^4\\
  		  & + 6.13\omega\theta^5 - 0.90\theta^6 - 46.15\omega^5 - 29.77\omega^4\theta - 58.22\omega^3\theta^2\\
  		  & - 54.43\omega^2\theta^3 - 21.75\omega\theta^4 - 34.48\theta^5 + 29.35\omega^4 + 1.80\omega^3\theta \\
		  &+ 58.86\omega^2\theta^2 - 20.33\omega\theta^3 + 42.83\theta^4  - 0.046\omega^3 - 0.01\omega^2\theta\\
		   &- 0.058\omega\theta^2  - 0.044\theta^3 + 1.09\times10^{-4}\omega^2 - 9.15\times10^{-5}\omega\theta\\\
		   &   + 2.046\times10^{-4}\theta^2
\else
	V=&-1.49\omega^6 + 2.45\omega^5\theta + 13.62\omega^4\theta^2 + 37.74\omega^3\theta^3 - 3.67\omega^2\theta^4 + 6.13\omega\theta^5 - 0.90\theta^6\\
	    & - 46.15\omega^5 - 29.77\omega^4\theta - 58.22\omega^3\theta^2 - 54.43\omega^2\theta^3 - 21.75\omega\theta^4 - 34.48\theta^5 + 29.35\omega^4\\
	    & + 1.80\omega^3\theta + 58.86\omega^2\theta^2 - 20.33\omega\theta^3 + 42.83\theta^4 - 0.046\omega^3 - 0.01\omega^2\theta - 0.058\omega\theta^2  \\ 
	    & - 0.044\theta^3 + 1.09\times10^{-4}\omega^2 - 9.15\times10^{-5}\omega\theta + 2.046\times10^{-4}\theta^2
\fi
\end{align*}}
\end{example}
The next example demonstrate how to employ the approach based on Bernstein polynomials on a multivariate nonlinearity.
\begin{example}\label{ex:bernstein}
Consider the system in~\eqref{eq:ex2:system}. This system can be approximated as a \nth{4} order polynomial as
\iftwocolumn\allowdisplaybreaks
\begin{align*}
	\dot x_1={}&x_2,\\
	\dot x_2={}&-2x_2-(- 9.5\times10^{-4}x_1^4x_2^3 + 0.015x_1^4x_2\\
	 &- 7.1\times10^{-4}x1^3x2^4 + 0.067x_1^3x_2^2 - 0.18x_1^3\\
	 & + 0.044x_1^2x_2^3- 0.7x_1^2x_2 + 3.6{\times}10^{-3}x_1x_2^4 - 0.34x_1x_2^2 \\
	 &+ 0.89x_1 + 3.7\times10^{-3}x_2^3 - 0.059x_2+\varepsilon)+u
\end{align*}
\else
\begin{align*}
	\dot x_1=&x_2,\\
	\dot x_2=&-2x_2-(- 9.5\times10^{-4}x_1^4x_2^3 + 0.015x_1^4x_2 - 7.1\times10^{-4}x1^3x2^4 + 0.067x_1^3x_2^2\\ 
	& - 0.18x_1^3 + 0.044x_1^2x_2^3 - 0.7x_1^2x_2 + 3.6\times10^{-3}x_1x_2^4 - 0.34x_1x_2^2 + 0.89x_1\\
	& + 3.7\times10^{-3}x_2^3  - 0.059x_2+\varepsilon)
\end{align*}
\fi
where \(\varepsilon\) is the approximation error and is bounded by \(-0.04\leq\varepsilon\leq0.04.\) \autoref{th:stabilityunforced3} proves that the system is locally stable for \(u=0\) based on the following \nth{4} order Lyapunov functional
\iftwocolumn
\begin{align*}
	V(\bx) ={}& 
  0.065802x_1^4 - 0.094308x_1^3x_2 - 0.036597x_1^2x_2^2\\
   &+ 0.0096327x_1x_2^3 + 0.0002283x_2^4 - 1.3876x_1^3\\
    &+ 0.037105x_1^2x_2 - 1.4013x_1   x_2^2 - 0.036844x_2^3\\
     &+ 2.0697x_1^2 + 0.33552x_1x_2 + 1.5356x_2^2
\end{align*}
\else
\begin{align*}
	V(\bx) ={}& 
  0.065802x_1^4 - 0.094308x_1^3x_2 - 0.036597x_1^2x_2^2 + 0.0096327x_1
x_2^3 + 0.0002283x_2^4\\
& - 1.3876x_1^3 + 0.037105x_1^2x_2 - 1.4013x_1
  x_2^2 - 0.036844x_2^3 + 2.0697x_1^2\\
  & + 0.33552x_1x_2 + 1.5356x_2^2,
\end{align*}
\fi
for \(-0.5\leq x_1,x_2\leq0.5.\) It can be shown that this system is also locally passive, using a \nth{6}-order Lyapunov function for \(\vert x_1\vert\leq0.5,\vert x_2\vert\leq0.5\) and \(\vert u\vert\leq0.5\). The Lyapunov function can be found using \autoref{th:bern:dissipativity} and \autoref{rem:passive}, however, it is not listed here for the sake of brevity. 
\end{example}

\section{Conclusions}\label{sec:conc}
In this paper, we proposed an optimisation-based approach to study certain energy-related behaviours of a nonlinear system through polynomial approximations. The behaviours of interest included stability, dissipativity, and passivity, charactrized by passivity indices. A motivating example was given to show that dissipativity and passivity of a system should be studied locally. Therefore, the focus here was on local properties of the system in well-defined admissible control and state spaces. The methodologies facilitate the systematic search for Lyapunov functionals through polynomial approximations. Two different approaches approximate the system's dynamics with polynomial functions. The first approach was through the well-known Taylor's theorem. This approach resulted in large optimisation programs, so we showed how we could reduce the size of the optimisation problem by using a generalised $\mathcal S$-procedure and by bounding the approximation errors in an ellipsoid. The second approach was through a multivariate generalisation of Bernstein polynomials. Examples were given to demonstrate the effectiveness and applicability of each approach. We showed that the approach based on Taylor's theorem provides a more intuitive approximation and is easier to derive for different regions; however, it may lead to larger optimisation programs, and there is a trade-off between accuracy and computational complexity. The second approach resulted in smaller optimisation problems and fewer computational requirements to solve the program. 
\printbibliography
\appendix
\section*{Appendix}
\begin{theorem}[Stone-Weierstrass Theorem] Let \(X\) be a compact Hausdorff space and \(A\) be a subalgebra of \(C(X, \bR)\) containing a non-zero constant function. Then \(A\) is dense in \(C(X, \bR)\) if and only if it separates points~\cite{rudin.analysis}. 
\end{theorem} 

\begin{theorem}[Multivariate version of Taylor's theorem~{\cite{Apostol}}]\label{th:Taylor}
If \(f : \bR^n \to\bR\) is a \(k\) times differentiable function at a point \(\boldsymbol a\in\bR^n,\) then there exist \(R_\beta : \bR^n\to\bR\) such that
\begin{equation}\label{eq:taylortheorem}
{\begin{aligned}&f({\bx })=\sum _{|\alpha |\leq k}{\frac {D^{\alpha }f({\boldsymbol {a}})}{\alpha !}}({\bx }-{\boldsymbol {a}})^{\alpha }+\sum _{|\beta |=k+1}R_{\beta }({\bx })({\bx }-{\boldsymbol {a}})^{\beta },\\&{\mbox{and}}\quad \lim _{{\bx }\to {\boldsymbol {a}}}R_{\beta }({\bx })=0.\end{aligned}}
\end{equation}
Here, the multi-index vectors \(\alpha\in\bR^n\) are the degrees of the monomials comprising the whole approximation and therefore, if \(\alpha=(\alpha_1,\dots,\alpha_n),\) then \(\bx^\alpha=x_1^{\alpha_1}\cdot x_2^{\alpha_2}\cdots x_n^{\alpha_n}.\) Also \(\vert\alpha\vert=\sum_{i=1}^n\alpha_i,\)  the derivative symbol in~\eqref{eq:taylortheorem} is defined as 
\begin{equation}
D^\alpha f(x)=\frac{\partial^{\vert\alpha\vert} f(x)}{\partial x_1^{\alpha_1}\cdots\partial x_n^{\alpha_n}},
\end{equation}
and \(\alpha!=\alpha_1!\alpha_2!\cdots\alpha_n!. \)
\end{theorem}

If the function \(f : \bR^n\to \bR\) is \(k + 1\) times continuously differentiable in the closed ball \(B\), then we can derive the remainder in terms of \((k+1)\)-th order partial derivatives of \(f\) in this neighborhood:
\[
{ {\begin{aligned}&f({\bx })=\sum _{|\alpha |\leq k}{\frac {D^{\alpha }f({\boldsymbol {a}})}{\alpha !}}({\bx }-{\boldsymbol {a}})^{\alpha }+\sum _{|\beta |=k+1}R_{\beta }({\bx })({\bx }-{\boldsymbol {a}})^{\beta },\\&R_{\beta }({\bx })={\frac {|\beta |}{\beta !}}\int _{0}^{1}(1-t)^{|\beta |-1}D^{\beta }f{\big (}{\boldsymbol {a}}+t({\bx }-{\boldsymbol {a}}){\big )}\,dt.\end{aligned}}}\] 
Here, based on the continuity of \((k+1)\)-th order partial derivatives in the compact set \(B\), we can obtain the uniform estimates
\begin{equation}\label{eq:remainderbound}
{\displaystyle \left|R_{\beta }({\bx })\right|\leq {\frac {1}{\beta !}}\max _{|\alpha |=|\beta |}\max _{{\boldsymbol {y}}\in B}|D^{\alpha }f({\boldsymbol {y}})|,\qquad {\bx }\in B.} 
\end{equation}

A Bernstein polynomial is a linear combination of Bernstein basis polynomials. For the univariate case, the \(m+1\) \emph{Bernstein basis polynomials} of degree \(m\) are defined as follows~\cite{lorentz_bernstein_1986}
\begin{equation}
b_{\nu,m}(x)=\binom m\nu  x^\nu(1-x)^{m-\nu},\quad \nu=0,\dots,m.
\end{equation}
The multivariate case can be defined similarly. 

\begin{definition}[Multivariate Bernstein Polynomials~{\cite{feng_asymptotic_1992}}]  Let  $ m_1,\ldots,m_n\in\mathbb{N}$ and $ f$ be a function of $ n$ variables. The polynomials
\iftwocolumn
\begin{multline}
 B_{m_1,\ldots,m_n}(f)(x_1,\ldots,x_n):=
\sum_{\stackrel{0\le k_j\le m_j}{1\leq j\leq n}}f\left(\frac{k_1}{m_1},\dots,\frac{k_n}{m_n}\right)\\
\times\prod_{j=1}^n \left( \binom{m_j}{k_j} x_j^{k_j} (1-x_j)^{m_j-k_j} \right)
\end{multline}
\else
\begin{equation}
 B_{m_1,\ldots,m_n}(f)(x_1,\ldots,x_n):=
\sum_{\stackrel{0\le k_j\le m_j}{1\leq j\leq n}}f\left(\frac{k_1}{m_1},\dots,\frac{k_n}{m_n}\right)
\prod_{j=1}^n \left( \binom{m_j}{k_j} x_j^{k_j} (1-x_j)^{m_j-k_j} \right)
\end{equation}
\fi
are called the multivariate Bernstein polynomials of $ f$.
We note that  $ B_{m_1,\ldots,m_n}(f)(\cdot)$ is a linear operator.
\end{definition}

The Bernstein polynomials of degree \(m\) are a basis for the vector space of polynomials of degree \(m\) or lower. 
A Bernstein polynomial is a linear combination of Bernstein basis polynomials
\begin{equation}
	B_m(x)=\sum_{\nu=0}^m\beta_m b_{\nu,m}(x).
\end{equation}
It is also called a polynomial in Bernstein form of degree \(m.\) 
\enlargethispage{\baselineskip}
\begin{theorem}
Consider a continuous function  \(f\) on the interval \([0, 1]\) and the Bernstein polynomial
\begin{equation}
{ B_{m}(f)(x)=\sum _{\nu =0}^{m}f\left({\frac {\nu }{m}}\right)b_{\nu ,m}(x).} 
\end{equation}
It can be shown that
\begin{equation}
\lim _{{m\to \infty }}{B_{m}(f)(x)}=f(x).
\end{equation}
The limit holds uniformly on the interval \([0, 1].\) This statement is stronger than pointwise convergence (where the limit holds for each value of x separately). Specifically, uniform convergence signifies that
\begin{equation}
\lim _{{m\to \infty }}\sup \left\{\,\left|f(x)-B_{m}(f)(x)\right|\,:\,0\leq x\leq 1\,\right\}=0.	
\end{equation}
\end{theorem}

\begin{theorem}[Uniform Convergence]   Let  $ f: [0,1]^n\to\bR$ be a continuous function. Then the multivariate Bernstein polynomials  $ B_{m_1,\ldots,m_n}(f)(\cdot)$ converge uniformly to $ f$ for  $ m_1,\ldots,m_n\to\infty.$ In other words, The set of all polynomials is dense in  $ C([0,1]^n).$
\end{theorem}
By assuming more knowledge about the function, specifically a Lipschitz condition, an error bound can be obtained.
\begin{theorem}[Error Bound for Lipschitz Condition]
If  $ f: [0,1]^n\to\mathbb{R}$ is a continuous function satisfying the Lipschitz condition
\begin{equation}
 \Vert f(x)-f(y)\Vert _2 < L \Vert x-y\Vert _2\vspace{-2mm}\vspace{-1mm}
\end{equation}
on $ [0,1]^n$, then the inequality
\begin{equation}\label{eq:bernstein.errorbound}\vspace{-4mm}\vspace{-1mm}
\Vert B_{m_1,\ldots,m_n}(f)(x) - f(x) \Vert _2 <
\frac L2 \biggl( \sum_{j=1}^n \frac1{m_j} \biggr)^\frac12
\end{equation}
holds.
\end{theorem}   

The following asymptotic formula gives us information about the rate of convergence.
\begin{theorem}[Asymptotic Formula]   Let  $ f: [0,1]^n\to\mathbb{R}$ be a $ C^2$ function and $ x\in [0,1]^n$, then
\iftwocolumn
\begin{gather}\notag
\lim_{m\to\infty} m ( B_{m,\ldots,m}(f)(x) - f(x) )
= \sum_{j=1}^n\frac{x_j(1-x_j)}{2}\frac{\partial^2f(x)}{\partial x_j^2}\\
\leq \frac18\sum_{j=1}^n \frac{\partial ^2 f(x)}{\partial x_j^2}.
\end{gather}
\else
\begin{equation}
\lim_{m\to\infty} m ( B_{m,\ldots,m}(f)(x) - f(x) )
= \sum_{j=1}^n\frac{x_j(1-x_j)}{2}\frac{\partial^2f(x)}{\partial x_j^2}
\leq \frac18\sum_{j=1}^n \frac{\partial ^2 f(x)}{\partial x_j^2}.
\end{equation}
\fi
\end{theorem}
The asymptotic formula states that the rate of convergence depends only on the partial derivatives  $ \partial ^2 f(x)/\partial x_j^2$. This is noteworthy, since it is often the case that the smoother a function is and the more is known about its higher derivatives, the more properties can be proven, but in this case only the second order derivatives play a role.

\enlargethispage{3mm}
The following theorem plays an important role in set inclusion results of polynomial nonnegativity. It is a simplified, and more tractable version of a well-known theorem called \emph{Positivstellensatz~\cite{parrilo_semidefinite_2003}.}
\begin{theorem}[Generalized $\mathcal S$-Procedure (See~\autocite{zakeri.2014,zakeri.2011} and the references therein)]
  Given polynomials \(\{p_i\}_{i=0}^m\subset\mathcal R_n,\) if there exists \(\{s_i\}_{i=1}^m\subset\Sigma_n\) such that\vspace{-4mm}\vspace{-1mm}
  \begin{equation}
    p_0-\sum_{i=1}^ms_ip_i\in\Sigma_n\vspace{-3mm}\vspace{-1mm}
  \end{equation}
  then \vspace{-1mm}\vspace{-1mm}
  \begin{equation}
    \cap\left\{\bx\in\bR^n\mid p_i(\bx)\geq0\right\}\subseteq
    \left\{\bx\in\bR^n\mid p_0(\bx)\geq0\right\}.
  \end{equation}\vspace{-1mm}\vspace{-1mm}
  Or equivalently, the following set is empty
  \begin{equation}
    \left\{\bx\in\bR^n\mid p_1(\bx)\geq0,\dots,p_m(\bx)\geq0,-p_0(\bx)>0\right\}
  \end{equation}
\end{theorem}

\end{document}